\documentclass[11pt, parskip=half, a4paper]{scrartcl}

\usepackage[english]{babel}		% Englisch
\usepackage{graphicx, subcaption, tikz}
\usepackage{listings}
\usepackage[hidelinks]{hyperref}
\usetikzlibrary{arrows.meta,bending,positioning}

\usepackage{amsmath, amssymb, amsbsy, amsthm}

\newcommand{\R}{\mathbb{R}}

\renewcommand{\phi}{\varphi}

\DeclareMathOperator{\tr}{tr}

\newcommand{\lb}{\left(}
\newcommand{\lsb}{\left[}
\newcommand{\lcb}{\left\{}

\newcommand{\rb}{\right)}
\newcommand{\rsb}{\right]}
\newcommand{\rcb}{\right\}}

\newcommand{\lpm}{\begin{pmatrix}}
\newcommand{\rpm}{\end{pmatrix}}

\usepackage{subcaption}
\usepackage{float}
\usepackage{amsmath}

\newtheorem{thm}{Theorem}[section]
\newtheorem{lem}[thm]{Lemma}

\newtheorem{rem}[thm]{Remark}

\title{Coupling opinion dynamics and epidemiology}
\subtitle{Preprint for arXiv}

\author{Thomas Goetz%
	\thanks{University of Koblenz, Germany, goetz@uni-koblenz.de}
	\and
	Tyll Krueger%
	\thanks{Wroclaw University of Science and Technology, Poland, tyll.krueger@pwr.edu.pl}
	\and
	Karol Niedzielewski
	\thanks{Interdisciplinary Centre for Mathematical and Computational Modelling (ICM), University of Warsaw, Poland, k.niedzielewski@icm.edu.pl}
	\and
	Jan Schneider
	\thanks{Wroclaw University of Science and Technology, Poland, jan.schneider@pwr.edu.pl}
    \and
    Barbara Pabjan
    \thanks{Wroclaw University of Science and Technology and Wroclaw University, Poland, barbara.pabjan@pwr.edu.pl}
	}

\date{\today\\ File:~\jobname}

\begin{document}
\maketitle
%\tableofcontents

\begin{abstract}

This research investigates the coupled dynamics of behavior and infectious disease using a mathematical model. We integrate a two-state q-voter opinion process with SIS-type infection dynamics, where transmission rates are influenced by the opinion and an infection-induced switching mechanism represents individuals reassessing their behavior upon infection. Analytically, we derive conditions for the stability of endemic and disease-free equilibria. Numerical simulations reveal complex dynamics: above a certain infectivity threshold, the system can exhibit alternative basins of attraction leading to a balanced endemic fixed point or stable limit cycles. Notably, the dominant asymptotic opinion and resulting epidemiological outcomes show non-monotonic relationships with infectivity, highlighting the potential for adaptive behavior to induce complex system dynamics. These findings underscore the critical role of social interventions; shifts in behavioral norms and trust can permanently alter epidemic outcomes, suggesting that such interventions are as crucial as biomedical controls.

\end{abstract}

\section{Introduction}

Behavior and infectious disease co-evolve. Protective practices—masking, physical distancing, ventilation, and vaccination—diffuse socially, while the evolving mix of protective and risky behaviors feeds back into epidemic trajectories. Even with biological parameters held fixed, coupled opinion–behavior dynamics can generate multiple asymptotic regimes (disease-free vs. endemic), tipping points, path dependence, and oscillations driven by endogenous behavior change. These features are well documented in the modeling literature that couples epidemiological processes with adaptive or strategic human responses  \cite{{Funk2010JRSI},{Fenichel2011PNAS},{Reluga2010PLoSCB}, {10.1007/978-3-031-61312-8_9}, {niedzielewski_chaos_2024}, {reitenbach_coupled_2024}}.
Because protective practices often require social reinforcement, their diffusion is well captured by complex-contagion mechanisms rather than purely independent adoption, as established in classic sociological and experimental network studies \cite{{CentolaMacy2007AJS},{Centola2010Science},{Ugander2012PNAS}}.

In this article we propose interpretations of parameter regimes and trajectories in ways that can be qualitative informative for pandemic management—specifically, for interventions that target compliance, trust, social learning, and behavioral spillovers across groups. Related work shows that altering the social process (e.g.~information diffusion, incentives, local imitation) can shift epidemic thresholds or even induce bistability and cycles; hence social policy can be as decisive as biomedical control   \cite{{Perra2011PLOSOne}, {Poletti2009JTB}, {Cornforth2011PLoSCB}}.

We study a tractable instance of an opinion–epidemic system by coupling a two-state $q$-voter opinion process (mutually exclusive states $A$ and $B$) with SIS-type infection dynamics whose transmission rates depend on the current opinion mix. In addition to distinct susceptibility and infectivity under $A$ vs. $B$, we assume that becoming infected induces a strong tendency to switch opinions (interpreted as individuals inferring that their prior behavior was insufficiently protective)—a mechanism akin to awareness- or experience-driven switching studied in behavior–disease models \cite{{Castellano2009PRE}, {Funk2010JTB}}.

As is common in epidemic modeling our baseline model assumes homogeneous mixing and constant recovery rates, omitting demography and seasonal forcing

Throughout the paper, state $B$ denotes the more protective behavioral repertoire in terms of infectivity (lower infectivity specified by parameter $\alpha$) and $A$ the less protective one. Additionally we assume that the states $A$ and $B$ can have a different susceptibility for getting infected (specified by the parameter $\delta$)

We contribute (i) an infection-induced switching mechanism within a two-state opinion process; (ii) analytical conditions for thresholds of disease-free states; and (iii) a characterization of non-monotonic long-run opinion/epidemic outcomes, including stable limit cycles.

Analytically and numerically, we show that (i) the fully-$A$ and fully-$B$ disease-free states have different stability regimes ; (ii) above an infectivity threshold, a balanced endemic fixed point exists exists together with additional endemic equilibria whose basins of attraction are separated by two dimensional unstable manifold of the balanced equilibria; and (iii) for certain parameter ranges, a stable limit cycle emerges (see Figures~\ref{F:Orbit} and~\ref{F:Orbit-zoom}). Notably, the dominant asymptotic opinion and the resulting epidemiological regime vary non-monotonically with infectivity --- a pattern consistent with prior demonstrations that adaptive behavior, imitation, and strategic responses can create backward bifurcations, multiple attractors, and long-run cycles in coupled systems. These properties matter for health policy: social interventions that shift the population across opinion basins can permanently change outcomes, while temporary suppression without altering the opinion mix risks rebound as parameters drift \cite{{Fenichel2011PNAS}, {Perra2011PLOSOne}, {Cornforth2011PLoSCB}, {BauchEarn2004PNAS}}.

\section{Model}

\subsection{Opinion dynamics}

Consider a population where individuals can have two mutually exclusive opinions $A$ and $B$; let $x$ denote the fraction of the population having opinion $A$ and $y=1-x$ denote the remaining part of the population having the other opinion $B$. Individuals may change their opinion when meeting with supporters of the opposite opinion; let $p$ denote the rate of change from opinion $A$ to $B$ or vice versa. For the sake of simplicity, we assume symmetry in both opinions. Then, we can set up a system of differential equations modeling the dynamics of the population size $x$ via
\begin{subequations}
\begin{align}
	x' &= -p x y^q + p y x^q\;,  \\
    y' &= p x y^q - p y x^q\;, 
\end{align}
\end{subequations}
where the first term describes the loss due to a switch from opinion $A$ to $B$ and the second term denotes an increase due to individuals switching from $B$ to $A$. The exponent $q\ge 1$ in the reaction terms can be interpreted as the number of individuals of opposite opinion needed to convince someone to change the own opinion. Using $y=1-x$ we arrive at
\begin{equation}
	x' = p x (1-x) \lb x^{q-1} - (1-x)^{q-1}\rb =: f(x) \;.
	\label{E:Voter}
\end{equation}
For $q=1$, this model is trivial, i.e.~$x'=0$ and hence no change in population share for each opinion. This is obvious because the number of individuals switching form opinion $A$ to $B$ equals to the number of people switching their opinion in the opposite direction.

Note, that $f'(x)=p \lb q (1-x) x^{q-1} + q x (1-x)^{q-1} -x^q - (1-x)^q\rb$.

For $q>1$, we obtain three possible equilibria:
\begin{itemize}
\item The two extremal (polarized) equilibria $x=0$ and $x=1$. Since $f'(0)=f'(1)=-p<0$ both polarized equilibria are asymptotically stable.
\item The balanced equilibrium $x=\frac{1}{2}$, where $f'(1/2)=\frac{p(q-1)}{2^{q-1}}>0$. Hence the balanced equilibrium is unstable.
\end{itemize}
To summarize, the above so--called \emph{$q$--voter model} describes a ''the winner takes it all'' behavior; a slight imbalance in the initial distribution e.g.~favoring $A$ over $B$, leads in the long run to a population where all individuals follow opinion $A$. The initial minority opinion $B$ dies out.

\subsection{Coupling opinion and infection dynamics}

Next, consider an infection process within a population with two opinions $A$ and $B$ influencing an individual's behavior with respect to the transmission of the infection. Let $S_A$, $S_B$, $I_A$ and $I_B$ denote the fraction of susceptibles or infectious with opinion $A$ or $B$ within the population. Neglecting demographic effects, we assume $S_A+S_B+I_A+I_B=1$. In the subsequent model we couple two dynamical processes, the transmission of and recovery from infection and changes of opinion. 

\begin{minipage}[t]{.6\textwidth}
Susceptibles change their opinion according to a $2$--voter model with rate $\tilde{p}>0$. 
Infectious of opinion $A$ transmit the disease with rate $\beta>0$ and infectious of the alternative opinion $B$ transmit it at a lower rate $\alpha\cdot \beta$, where $\alpha\in [0,1]$ models the reduced infectivity associated with opinion $B$. In addition, susceptible of the alternative opinion $B$ have a modified risk of getting infected. The parameter $\delta>0$ describes, change of the infection risk. A value of $\delta<1$ implies, that opinion $B$ also \emph{reduces} the risk of getting infected, while $\delta>1$ models the interesting situation, that opinion $B$ \emph{increases} the risk of getting infected. An example for such a situation can be COVID--19 and the opinions about mask wearing; $A$ describes non--mask--wearers and $B$ are mask-wearers. In case of an infection, an individual always switches opinion, since the prior opinion did not prevent the infection. Recovery occurs at fixed rate $\gamma>0$, independent of the opinion.
\end{minipage}
\hfill
\begin{minipage}[t]{.35\textwidth}
\vspace*{0pt}\par
\begin{tikzpicture}[scale=1]
\draw[fill=green!80!black] (0,3) circle (0.5cm) node {$S_a$};
\draw[fill=green] (0,0) circle (0.5cm) node {$S_b$};
\draw[fill=orange!80!black] (3,3) circle (0.5cm) node {$I_a$};
\draw[fill=orange] (3,0) circle (0.5cm) node {$I_b$};
\draw[-Stealth, green!80!black, very thick] (2.5,3) to [out=135, in=45] 
	node[above, green!80!black]{$\gamma$} (0.5,3) ;
\draw[-Stealth, green!80!black, very thick] (2.5,0) to [out=135, in=45, bend left] 
	node[below, green!80!black]{\parbox{1cm}{$\quad \gamma$\newline recovery}} (0.5,0);
\draw[-Stealth, red, very thick] (0.45,2.8) to [out=135, in=135, bend left] (3,0.5) ;
\draw[-Stealth, red, very thick] (0.45,0.2) to [out=135, in=135, bend right] (3,2.5);
\node[red] at (1,0.6) {$\delta$};
\node[right, red] at (2.7,1.5) {\parbox{2cm}{infection\newline$\beta (I_a+\alpha I_b)$}};
\draw[-Stealth, black, very thick] (-0.3,2.6) to [out=225, in=135, bend right] 
	node[left]{$\tilde{p} S_b^2$} (-0.3,0.4);
\draw[-Stealth, black, very thick] (0.3,0.4) to [out=225, in=135, bend right] 
	node[right]{$\tilde{p} S_a^2$} (0.3,2.6);
\node[rotate=-90, black] at (0,1.5) {\parbox{1cm}{opinion change}};
\end{tikzpicture}

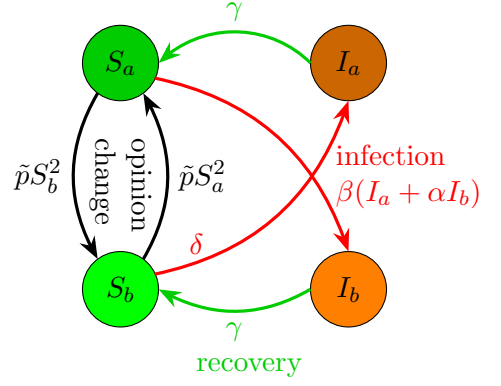
\captionof{figure}{\label{F:Sketch_DOS} Sketch of the transitions in the coupled disease--opinion system~\eqref{E:SIsystem}.}
\end{minipage}

This leads us to the following system
\begin{subequations}
\label{E:SIsystem}
\begin{align}
	S_a' &= -\beta S_a (I_a+\alpha I_b) + \gamma I_a + \tilde{p} S_a S_b (S_a-S_b)\;, \\
	S_b' &= -\beta\delta S_b (I_a+\alpha I_b) + \gamma I_b + \tilde{p} S_a S_b (S_b-S_a)\;, \\
	I_a' &= \beta \delta S_b (I_a+\alpha I_b)-\gamma I_a\;, \\
	I_b' &= \beta S_a (I_a+ \alpha I_b) - \gamma I_b\;.
\end{align}
\end{subequations}

The feasible domain for the system~\eqref{E:SIsystem} is given by the unit simplex 
\begin{equation*}
    \Omega=\lcb 0\le S_a,S_b,I_a,I_b\le 1, \text{ and } S_a+S_b+I_a+I_b=1\rcb 
    \subset \R^4\;.
\end{equation*}

In the sequel, we will concentrate most of our analytical investigations to the case of $\delta=1$. For the general situation of $\delta>0$, we will show later on numerical results, see Section~\ref{Sec:numerical}.

For the sake of compact notation, we introduce $x=S_a$, $y=S_b$ and $v=I_a+\alpha I_b$. The fourth compartment can be eliminated using the constant population assumption. We scale time with the recovery period and set $R=\beta/\gamma$ as the basic reproduction number of the classical SIS--system as well as $p=\tilde{p}/\gamma$. In the $xyv$--variables and using $\delta=1$, system~\eqref{E:SIsystem} reads as
\begin{subequations}
\label{E:System:xyv}
\begin{align}
	x' &= -Rxv + pxy(x-y) + \frac{v-\alpha(1-x-y)}{1-\alpha} \label{E:System:xyv:x}\\
	y' &= -Ryv + pxy(y-x) + \frac{1-v-x-y}{1-\alpha} \label{E:System:xyv:y}\\
    v' &= R(\alpha x+y)v-v \label{E:System:xyv:v}
\intertext{and the Jacobian given by}
    J_{xyv} &= \lpm -Rv+py(2x-y)+\frac{\alpha}{1-\alpha} & px(x-2y)+\frac{\alpha}{1-\alpha} & -Rx+\frac{1}{1-\alpha} \\
	 py(y-2x)-\frac{1}{1-\alpha} & -Rv+px(2y-x)-\frac{1}{1-\alpha} & -Ry-\frac{1}{1-\alpha} \\
	 \alpha Rv & Rv & R(\alpha x+y)-1\rpm \label{E:Jac:xyv}
\end{align}
\end{subequations}
A special case of~\eqref{E:SIsystem} is obtained, if opinion B leads to no transmission of the disease, i.e.~considering the limiting case of $\alpha=0$. This case will be discussed later on in more detail.

Introducing as new coordinates the total number of susceptibles $s=x+y$ and $z=\alpha x+y$, we arrive again for the $(\delta=1)$--case at the alternative formulation
\begin{subequations}
\label{E:System:szv}
\begin{align}
    s' &= -Rsv + 1-s \label{E:System:szv:s}\\
    z' &= -Rzv + \frac{p}{(1-\alpha)^2} (s-z) (\alpha s-z) \lb (1+\alpha)s-2z \rb -v + (1+\alpha)(1-s) \label{E:System:szv:z}\\
    v' &= Rzv-v \label{E:System:szv:v}\;.
\intertext{In the $szv$--coordinates, the Jacobian reads as}
	J_{szv} &= \lpm -Rv-1 & 0 & -Rs \\
	 J_{21} & J_{22} & -Rz-1 \\
	 0 & Rv & Rz-1\rpm  \label{E:Jac:szv}
\end{align}
\end{subequations}
where 
\begin{align*}
    J_{21} &= \frac{p}{(1-\alpha)^2}\lsb 3(1+\alpha) (\alpha s^2+z^2)-2(1+4\alpha+\alpha^2)zs\rsb -(1+\alpha)
\intertext{and}
    J_{22} &= -Rv -\frac{p}{(1-\alpha)^2}\lsb 6z(z-(1+\alpha)s)+(1+4\alpha+\alpha^2)s^2\rsb\;.
\end{align*}
The feasible domain can be stated as
\begin{equation*}
    \Omega_{szv} = \lcb 0\le s\le 1, \text{ and } \alpha(1-s)\le v\le 1-s,
     \text{ and } \alpha s\le z\le s\rcb\subset\R^3\,.
\end{equation*}

\begin{rem}(see~\cite{LL98}) To determine, if a $3\times 3$-matrix $A$ is stable, i.e.~if all the three eigenvalues have negative real part, one has to check the following three conditions
\begin{enumerate}
    \item $\tr A < 0$,
    \item $\det A < 0$ and
    \item $\tr A \cdot S_2(A) < \det A$, where $S_2(A)$ denotes the sum of the principal $2\times 2$ minors given by
    \begin{equation*}
        S_2(A) = \det \lpm a_{11} & a_{12} \\ a_{21} & a_{22}\rpm +
                \det \lpm a_{11} & a_{13} \\ a_{31} & a_{32}\rpm +
                \det \lpm a_{22} & a_{23} \\ a_{32} & a_{33}\rpm\;.
        \end{equation*}
\end{enumerate}   
\end{rem}

%
%
% TG work in progress
\section{Analysis of the equilibria}
In this section we analyze the different equilibria of system~\eqref{E:System:xyv} and their stability. As usual, one can distinguish the \emph{disease free equilibria} characterized by $v=0$ from the \emph{endemic equilibria} $v>0$. 

First, let us focus on the disease free equilibria. Analogous to the pure $q$--voter model, the above system~\eqref{E:System:xyv} allows for three disease free equilibria given in $xyv$--coordinates by $\text{DFE}_0=(0,1,0)$, $\text{DFE}_1=(1,0,0)$ and $\text{DFE}_b=(\frac12, \frac12, 0)$. Inspecting the eigenvalues of the Jacobian for the disease free equilibria, we directly obtain

\begin{lem}\label{L:DFE:a}
For $\alpha\in [0,1]$ and $\delta=1$, the following holds
\begin{enumerate}
    \item The disease free equilibrium $\text{DFE}_0=(0,1,0)$ is asymptotically stable, if and only if $R<1$.
    \item The disease free equilibrium $\text{DFE}_1=(1,0,0)$ is asymptotically stable, if and only if $R<\frac{1}{\alpha}$.
    \item The balanced disease free equilibrium $\text{DFE}_b=(\frac12, \frac12, 0)$ is unstable for all $R>0$.
\end{enumerate}
\end{lem}

\begin{rem}\label{R:DFE:a0}
In the special case of $\alpha=0$, i.e.~infected individuals of opinion $B$ do \emph{not} contribute to the transmission of the disease, the disease free equilibrium $\text{DFE}_1$ is asymptotically stable \emph{for all} values of $R>0$.
\end{rem}

Next, we investigate endemic equilibria. It turns out to be advantageous to switch to $szv$--coordinates introduced in~\eqref{E:System:szv}. We easily derive $z=\frac{1}{R}$ and $v=\frac{1-s}{Rs}$ from~\eqref{E:System:szv:v} and~\eqref{E:System:szv:s}. Inserting into~\eqref{E:System:szv:z}, we obtain the following equation for $s$:
\begin{equation}
\label{E:Endemic:s}
    0 = \lsb (1+\alpha)s-\frac{2}{R}\rsb \cdot \lsb \frac{p}{(1-\alpha)^2} \lb s-\frac{1}{R}\rb \lb \alpha s-\frac{1}{R} \rb + \frac{1-s}{s}\rsb\;.
\end{equation}

The first factor yields the solution $s_b=\frac{2}{(1+\alpha)R}$, i.e.~$x_b=y_b=\frac{1}{(1+\alpha)R}$ and $v_b=\frac{(1+\alpha)R-2}{2R}$. This balanced endemic equilibrium $\text{EE}_b$ is feasible, provided $(1+\alpha) R>2$. The Jacobian at the balanced endemic equilibrium reads in the $szv$--formulation as
\begin{equation}
	J_b := J(\text{EE}_b) = \lpm - \frac{1+\alpha}{2}R & 0 & - \frac{2}{1+\alpha} \\
	    -(1+\alpha)(1+\rho) & 1-\frac{1+\alpha}{2}R+2\rho & -2 \\
	    0 & \frac{1+\alpha}{2}R-1 & 0 \rpm\;,  
\end{equation}
where $\rho=\frac{p}{(1+\alpha)^2 R^2}>0$. The three stability conditions can be translated into conditions for $p$ depending on $R$. Setting $Q=(1+\alpha) R$ for shorter notation we get
\begin{enumerate}
\item $\tr J_b = 1-Q+2\frac{p}{Q^2}<0$ 
    yields $p<\frac{Q^2}{2}\lb Q-1\rb$.
\item $\det J_b = \lb \frac{Q}{2}-1\rb \lb 2+2\frac{p}{Q^2}-Q\rb<0$ yields $p<\frac{Q^2}{2}\lb Q-2\rb$ because of $Q>2$. 
Note, that the condition $p<\frac{Q^2}{2}\lb Q-2\rb$ automatically implies the first condition $p<\frac{Q^2}{2}\lb Q-1\rb$.
\item The last condition involves $S_2(J_b)= \frac{Q^2}{4}+\frac{Q}{2}-\frac{p}{Q}-2$. Then the third condition reads as
\begin{equation*}
    \tr J_b\cdot S_2(J_b)-\det J_b = -\frac{\lb Q^3-4p\rb \cdot\lb Q^3-Q^2-2Q-2p \rb}{4Q^3}  < 0
    \end{equation*}
    Therefore either $p>\max\lcb \frac{Q^3}{4}, \frac{Q^3}{2}-\frac{Q^2}{2}-Q\rcb$ or $p<\min\lcb \frac{Q^3}{4}, \frac{Q^3}{2}-\frac{Q^2}{2}-Q\rcb$.
\end{enumerate}
Summing up all three conditions we readily obtain

\begin{lem}\label{L:EEb:a}
For $\alpha\in [0,1]$ and $\delta=1$, the balanced endemic equilibrium $\text{EE}_b$ with $xyv$--coordinates $(x_b,y_b,v_b)=\lb\frac{1}{(1+\alpha)R}, \frac{1}{(1+\alpha)R}, \frac{1+\alpha}{2}-\frac{1}{R}\rb$ exists, if $R>\frac{2}{1+\alpha}$. It is asymptotically stable, if 
    \begin{equation*}
        p < \frac{(1+\alpha)^2 R^2}{2}((1+\alpha)R - 2)
        \quad \textrm{and}\quad 
        p < \begin{cases}
             \frac{(1+\alpha)^3}{2}R^3-\frac{(1+\alpha)^2}{2} R^2 - (1+\alpha)R & \text{for } \frac{2}{1+\alpha}\le R\le \frac{4}{1+\alpha}\;,\\
             \frac{(1+\alpha)^3}{4}R^3 & \text{for } R> \frac{4}{1+\alpha}\;.
        \end{cases}
    \end{equation*}
\end{lem}

Other endemic equilibria are roots of the second factor in~\eqref{E:Endemic:s}. Here, the case $\alpha=0$ is considerably simpler than $\alpha>0$ because the order in $s$ reduces for $\alpha=0$. 

For $\alpha=0$, we have to solve the quadratic equation
\begin{equation*}
    s^2 + \lsb \frac{R}{p} - \frac{1}{R}\rsb s + \frac{R}{p} = 0
\end{equation*}
and obtain one positive solution $s^\ast$ and hence $x^\ast=s^\ast-\frac{1}{R}$ equals to
\begin{equation*}
    x^\ast = \frac{\sqrt{p^2+R^4+2pR^2(2R-1)}-p-R^2}{2pR}\;.
\end{equation*}
Using \eqref{E:System:xyv:x} we directly obtain the relation $v^\ast=\frac{p}{R^2}x^\ast$ for $\alpha=0$. At the according endemic equilibrium $E^\ast$ with $xyv$--coordinates given by $\lb x^\ast, \frac{1}{R}, \frac{p}{R^2}x^\ast\rb$, the Jacobian reads as
\begin{equation}
    J^\ast = J(E^\ast) = \lpm \frac{p}{R^2}(Rx^\ast-1) & \frac{p}{R}x^\ast (Rx^\ast-2) & 1-Rx^\ast \\
        \frac{p}{R^2} (1-2Rx^\ast)-1 & -p{x^\ast}^2+\frac{p}{R}x^\ast-1 & -2 \\ 0 & \frac{p}{R} x^\ast & 0\rpm\;. 
\end{equation}
It is immediate to check, that the trace of the Jacobian equals $\tr J = -\frac{p}{R^2} ( Rx^\ast-1)^2-1$ is always negative. For the determinant we obtain 
\begin{equation*}
    \det J^\ast = -\frac{p}{R} x^\ast \lb 1-Rx^\ast\rb \lb 2\frac{p}{R}x^\ast + \frac{p}{R^2}+1\rb\;.
\end{equation*}
Negativity of the determinant implies $x^\ast<\frac{1}{R}$, i.e.~$x^\ast$ is smaller than for the balanced endemic case. Substituting the above expression for $x^\ast$, one can derive, that $x^\ast<\frac{1}{R}$ is satisfied for all $p>0$ if $R\in (0,2]$ and for $p>\frac{1}{2}R^2(R-2)$ if $R>2$. 

\begin{figure}[htp]
\centerline{\includegraphics[width=.9\linewidth]{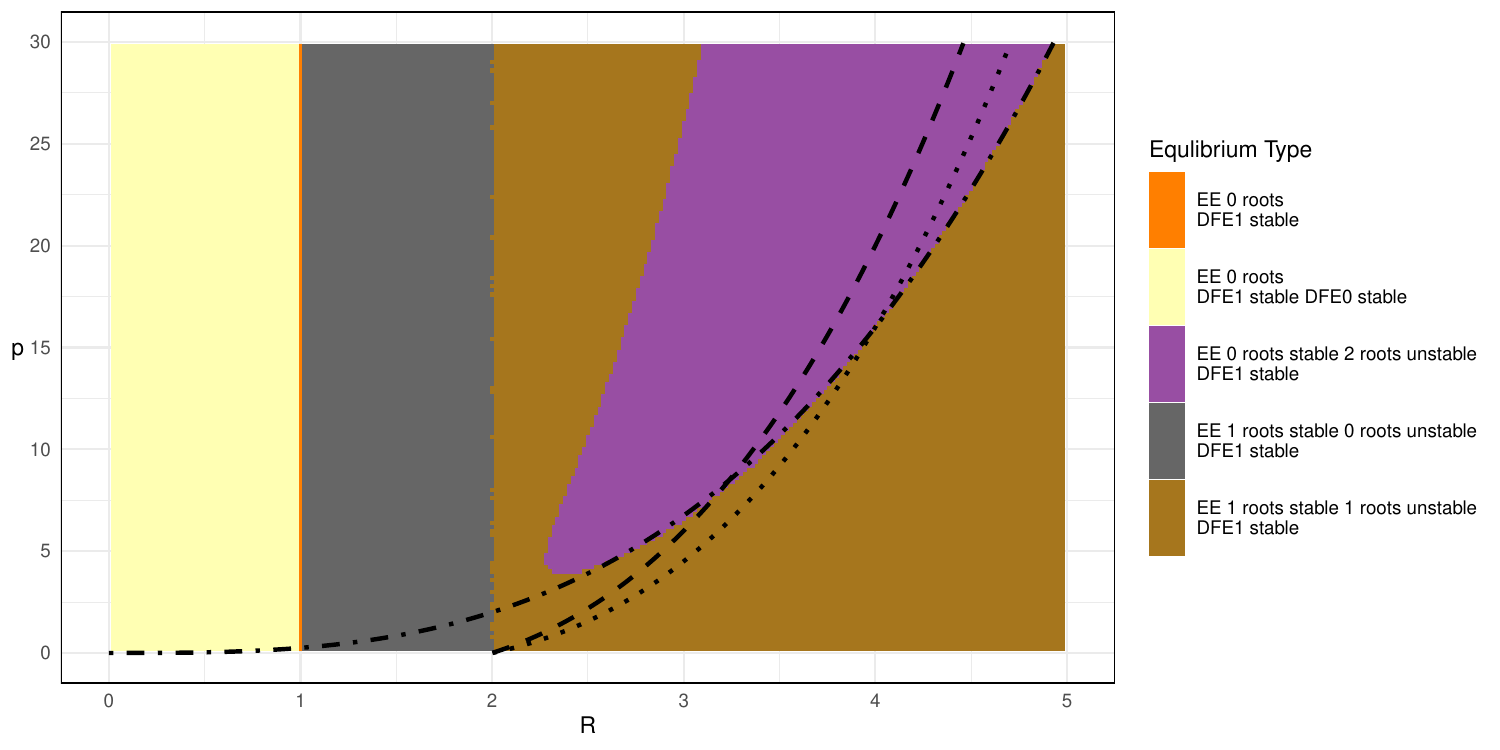}}
\caption{\label{F:heatmap:a0} Stability map in the $R$--$p$ plane of the model with $\alpha=0$. The colors indicate the stability domains of the different equilibria. The curves show the parameter bounds derived in Lemmata~\ref{L:EEb:a}; $\det A < 0$ formula is represented by the dotted line, $\tr A \cdot S_2(A) < \det A$ formula is represented by the dashed line and dot--dashed line. The acronyms in the legend translate to Endemic Equilibrium (EE) and Disease Free Equilibrium 1 (DFE$_1$) from Lemmata~\ref{L:DFE:a}. The orange region is visible as a narrow vertical line with $R=1.0$}
\end{figure}

The third stability condition $\tr J^\ast\cdot S_2(J^\ast)-\det J^\ast<0$ renders a complicated fifth order polynomial inequality in $R$, $p$ and $x^\ast$. Because we lack an easy interpretation of the terms involved, we skip the analytic details. We rather present a heat map in the $Rp$--plane depicting the stability region of the different equilibria, see Fig.~\ref{F:heatmap:a0}. As stated in Remark~\ref{R:DFE:a0}, the disease free equilibrium $\text{DFE}_1$ is always stable. In the light yellow region, where $R<1$, the other disease free equilibria $\text{DFE}_0$ is stable as well. Inside the grey region $1< R< 2$, the endemic equilibrium $E^\ast$ appears to be stable. For the brown domain, only one of the two endemic equilibria $\text{EE}_b$ and $E^\ast$ is stable, while in the red purple, both endemic equilibria get unstable. The dotted curve $p=R^3/2-R^2$, the dash--dotted curve $p=R^3/4$ and the dashed curve $p=R^3/2-R^2/2-R$ show the $p$--bounds derived in Lemma~\ref{L:EEb:a}. The dash--dotted curve and the dashed curves intersect at $R=4$. Below the minimum of three curves, $\text{EE}_b$ is stable.

Now, let us consider other endemic equilibria in the more general case $0<\alpha<1$. The second factor of Eqn.~\eqref{E:Endemic:s} can be rewritten either as the intersection of a hyperbola $h$ with a parabola $g_p$
\begin{gather}
    h(s) := \frac{1}{s}-1 \stackrel{!}{=} \frac{\alpha p}{(1-\alpha)^2} \lb s-\frac{1}{R}\rb \lb \frac{1}{\alpha R}-s\rb =: g_p(s) \label{E:equi:s}
\intertext{or as a single cubic equation}
    0 = \chi(s) := \alpha q R^3 s^3 - (1+\alpha) q R^2 s^2 + (q R-1) s +1 \label{E:cubic}\;,
\end{gather}
where $q=\frac{p}{(1-\alpha)^2R^3}$.

The hyperbola has one root at $s=1$ and the parabola $g_p$ has two roots at $1/R$ and $1/(\alpha R)$. We remark, that the balanced equilibrium $s_b=\frac{2}{R(1+\alpha)}$ is the harmonic mean of these two roots. 

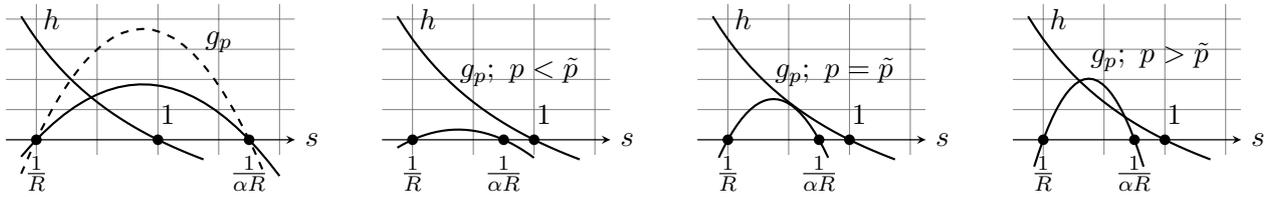
\begin{figure}[hbt]
\begin{tikzpicture}[xscale=4, yscale=2]
\draw [step=0.2,very thin,color=gray] (0.5,-0.1) grid (1.45,0.9); % Gitter zeichnen
    \draw[-stealth] (0.5,0) -- (1.45,0) node[right] {$s$}; % Achsen zeichnen
    \draw[domain=0.55:1.15, thick] plot (\x, {1/\x-1});
    \node at (0.65, .8) {$h$}; 
    \draw[domain=0.55:1.4, thick] plot (\x, {3*(\x-.6)*(1.3-\x)});
    \draw[domain=0.55:1.35, dashed, thick] plot (\x, {6*(\x-.6)*(1.3-\x)});
    \node at (1.2,0.65) {$g_p$};
    \node[circle,fill, inner sep=0pt,minimum size=4pt, label=below:$\tfrac{1}{R}$] at (0.6,0) {};
    \node[circle,fill, inner sep=0pt,minimum size=4pt, label=above:$\ \ 1$] at (1,0) {};
    \node[circle,fill, inner sep=0pt,minimum size=4pt, label=below:$\tfrac{1}{\alpha R}$] at (1.3,0) {};
\end{tikzpicture}
\hfill
\begin{tikzpicture}[xscale=4, yscale=2]
\draw [step=0.2,very thin,color=gray] (0.5,-0.1) grid (1.25,0.9); % Gitter zeichnen
    \draw[-stealth] (0.5,0) -- (1.25,0) node[right] {$s$}; % Achsen zeichnen
    \draw[domain=0.55:1.15, thick] plot (\x, {1/\x-1});
    \node at (0.65, .8) {$h$}; 
    \draw[domain=0.55:1, thick] plot (\x, {3*(\x-.6)*(.9-\x)});
    \node at (.95,0.45) {$g_p;\ p<\tilde{p}$};
    \node[circle,fill, inner sep=0pt,minimum size=4pt, label=below:$\tfrac{1}{R}$] at (0.6,0) {};
    \node[circle,fill, inner sep=0pt,minimum size=4pt, label=above:$\ \ 1$] at (1,0) {};
    \node[circle,fill, inner sep=0pt,minimum size=4pt, label=below:$\tfrac{1}{\alpha R}$] at (.9,0) {};
\end{tikzpicture}
\hfill
\begin{tikzpicture}[xscale=4, yscale=2]
\draw [step=0.2,very thin,color=gray] (0.5,-0.1) grid (1.25,0.9); % Gitter zeichnen
    \draw[-stealth] (0.5,0) -- (1.25,0) node[right] {$s$}; % Achsen zeichnen
    \draw[domain=0.55:1.15, thick] plot (\x, {1/\x-1});
    \node at (0.65, .8) {$h$}; 
    \draw[domain=0.57:.93, thick] plot (\x, {12*(\x-.6)*(.9-\x)});
    \node at (.95,0.45) {$g_p;\ p=\tilde{p}$};
    \node[circle,fill, inner sep=0pt,minimum size=4pt, label=below:$\tfrac{1}{R}$] at (0.6,0) {};
    \node[circle,fill, inner sep=0pt,minimum size=4pt, label=above:$\ \ 1$] at (1,0) {};
    \node[circle,fill, inner sep=0pt,minimum size=4pt, label=below:$\tfrac{1}{\alpha R}$] at (.9,0) {};
\end{tikzpicture}
\hfill
\begin{tikzpicture}[xscale=4, yscale=2]
\draw [step=0.2,very thin,color=gray] (0.5,-0.1) grid (1.25,0.9); % Gitter zeichnen
    \draw[-stealth] (0.5,0) -- (1.25,0) node[right] {$s$}; % Achsen zeichnen
    \draw[domain=0.55:1.15, thick] plot (\x, {1/\x-1});
    \node at (0.65, .8) {$h$}; 
    \draw[domain=0.57:.93, thick] plot (\x, {18*(\x-.6)*(.9-\x)});
    \node at (.95,0.55) {$g_p;\ p>\tilde{p}$};
    \node[circle,fill, inner sep=0pt,minimum size=4pt, label=below:$\tfrac{1}{R}$] at (0.6,0) {};
    \node[circle,fill, inner sep=0pt,minimum size=4pt, label=above:$\ \ 1$] at (1,0) {};
    \node[circle,fill, inner sep=0pt,minimum size=4pt, label=below:$\tfrac{1}{\alpha R}$] at (.9,0) {};
\end{tikzpicture}
\caption{\label{F:Sketch_s} Sketch of the equation~\eqref{E:equi:s} determining the endemic equilibria. Both, the hyperbola $h(S)=\frac{1}{s}-1$ and the parabola $g_p(s)$ are shown. Depending on the location of the roots $1/R$ and $1/(\alpha R)$ and the opinion change rate $p$, we get zero, one or two solutions for $s$ in the interval $[0,1]$.}
\end{figure}

If $1<R<1/\alpha$, then there exists only one solution $s_1\in (1/R,1)$ of Eqn.~\eqref{E:equi:s}, see Fig.~\ref{F:Sketch_s}(left). 

For $R>1/\alpha$, the number of solutions depends on the opinion change parameter $p$. For a critical value $p=\tilde{p}(R,\alpha)$, there is exactly one intersection of the hyperbola and the parabola as shown in Fig.~\ref{F:Sketch_s}(second from right), for $p<\tilde{p}$ there is no intersection (Fig.~\ref{F:Sketch_s}(second from left)) and for $p>\tilde{p}$ we get two intersections in the relevant domain $0\le s\le 1$, see Fig.~\ref{F:Sketch_s}(right).

The critical threshold $\tilde{p}$, where two additional endemic equilibria are appearing, can be determined by inspecting the discriminant of the cubic equation $\chi$. However, the resulting expressing involves higher power of $R$, $\alpha$ and $p$ or $q$ that do not allows us an easy interpretation. 

\begin{figure}[H]
    \centering
    \begin{subfigure}[t]{\textwidth}
        \centering
        \includegraphics[width=.9\textwidth]{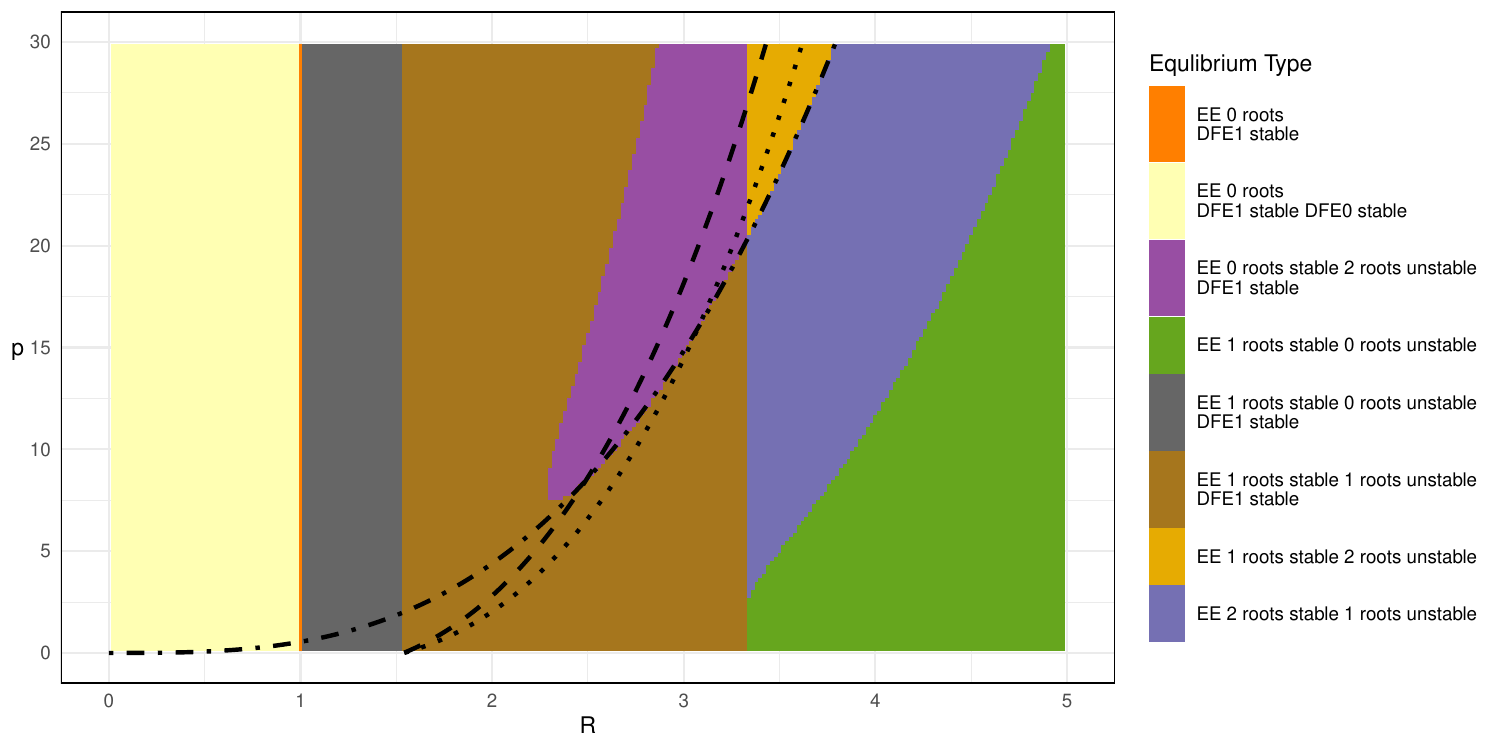}
        \caption{$\alpha = 0.3$}\label{subfig:Heatmap2_a}
    \end{subfigure}%
    \vspace{15mm}
    \begin{subfigure}[t]{\textwidth}
        \centering
        \includegraphics[width=.9\textwidth]{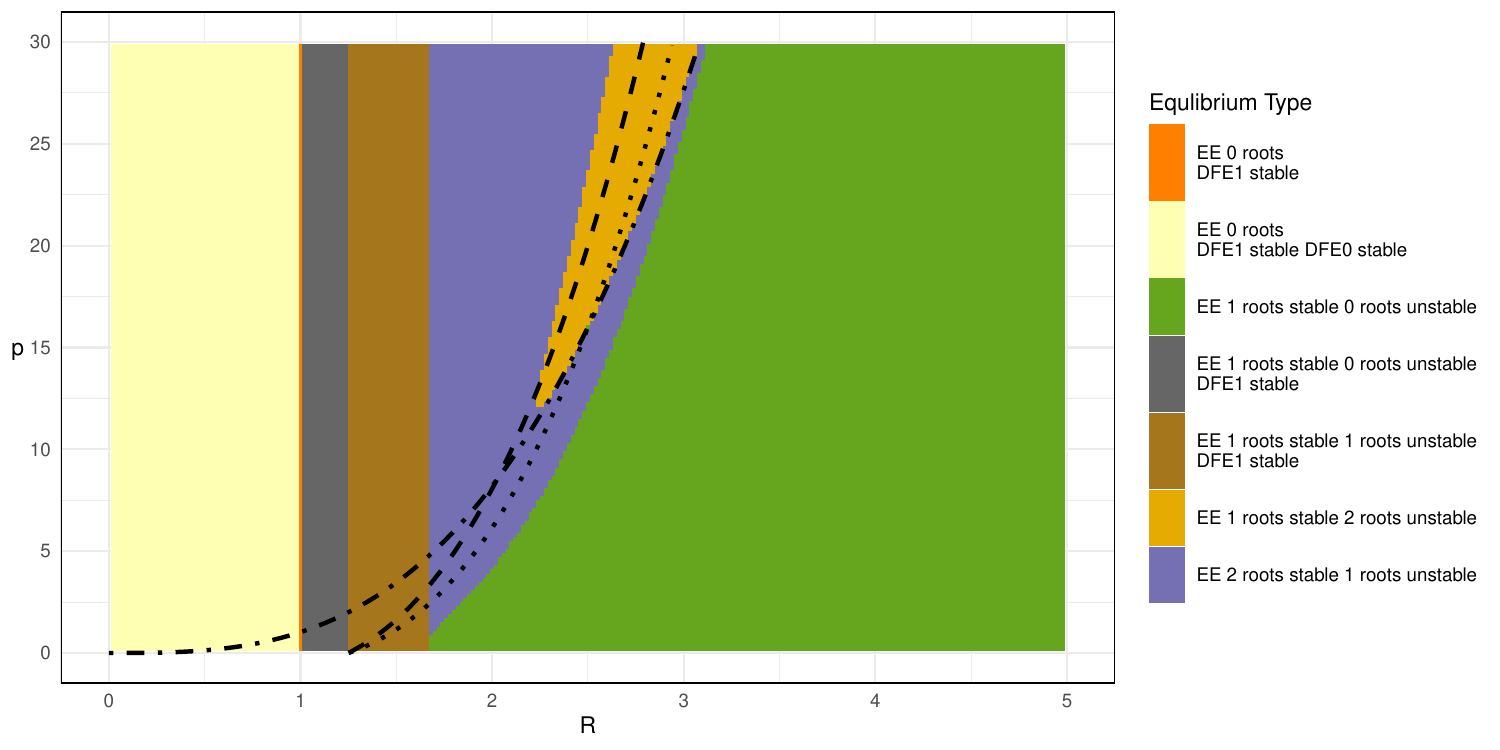}
        \caption{$\alpha = 0.6$}\label{subfig:Heatmap2_b}
    \end{subfigure}
    \caption{Stability domains in the $R$--$p$ plane for $\alpha=0.3$ (upper figure) and $\alpha=0.6$ (bottom figure). The colors indicate the stability domains of the different equilibria. The curves show the parameter bounds derived in Lemmata~\ref{L:EEb:a}; $\det A < 0$ formula is represented by the dotted line, $\tr A \cdot S_2(A) < \det A$ formula is represented by the dashed line and dot--dashed line. The acronyms in the legend translate to Endemic Equilibrium (EE), Disease Free Equilibrium 1 (DFE$_1$) and Disease Free Equilibrium 2 (DFE$_2$) from Lemmata~\ref{L:DFE:a}. Orange region is visible as a narrow vertical line with $R=1.0$. We can see very different characteristics of dynamics between figures. Colors mostly overlap, however their sizes alter greatly. Purple region ($DFE_1$ stable) disappears when going from $\alpha=0.3$ to $\alpha=0.6$.}\label{F:Heatmap2}
\end{figure}

\begin{figure}[H]
    \centering
    \begin{subfigure}[t]{\textwidth}
        \centering
        \includegraphics[width=.9\textwidth]{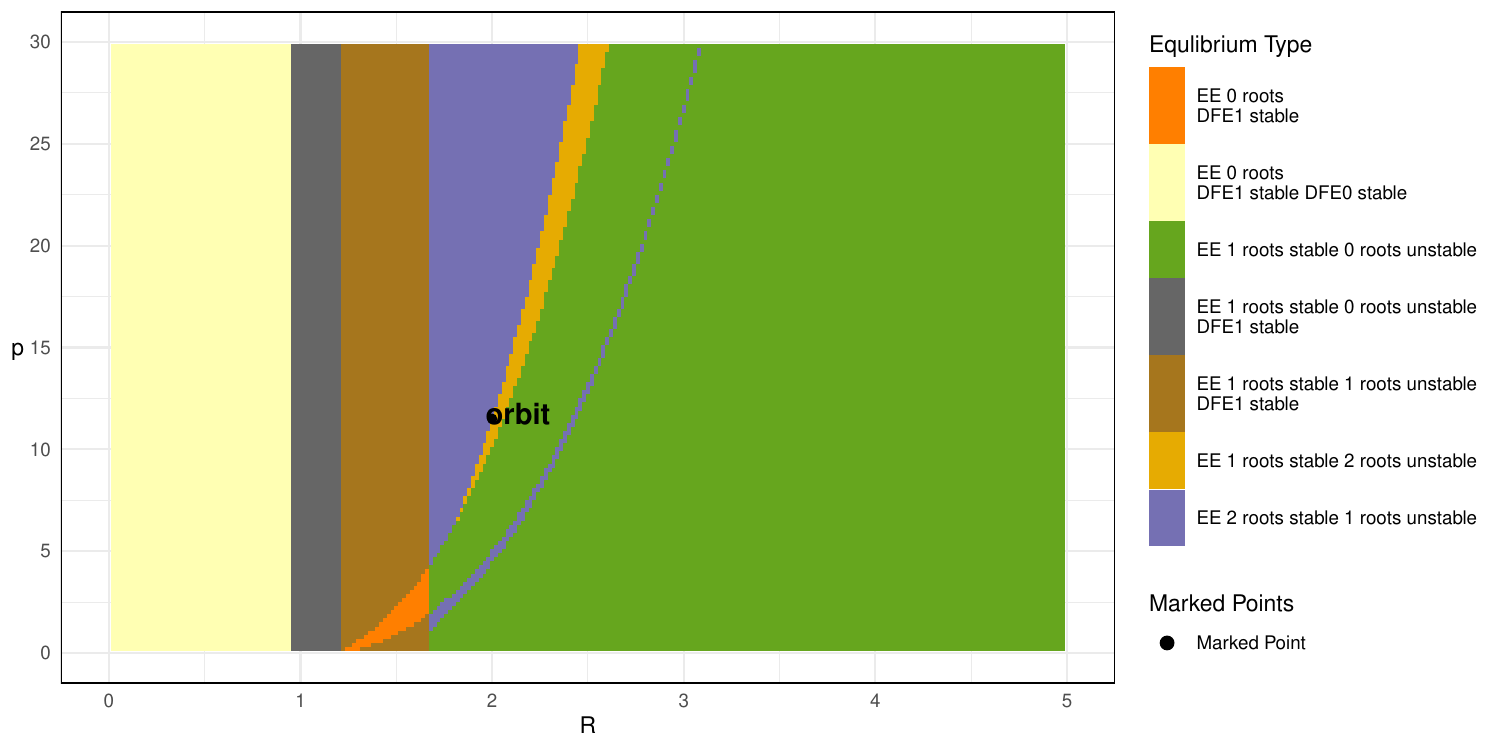}
        \caption{$\delta = 1.0530242665286913$}\label{subfig:Heatmap3_a}
    \end{subfigure}%
    \vspace{15mm}
    \begin{subfigure}[t]{\textwidth}
        \centering
        \includegraphics[width=.9\textwidth]{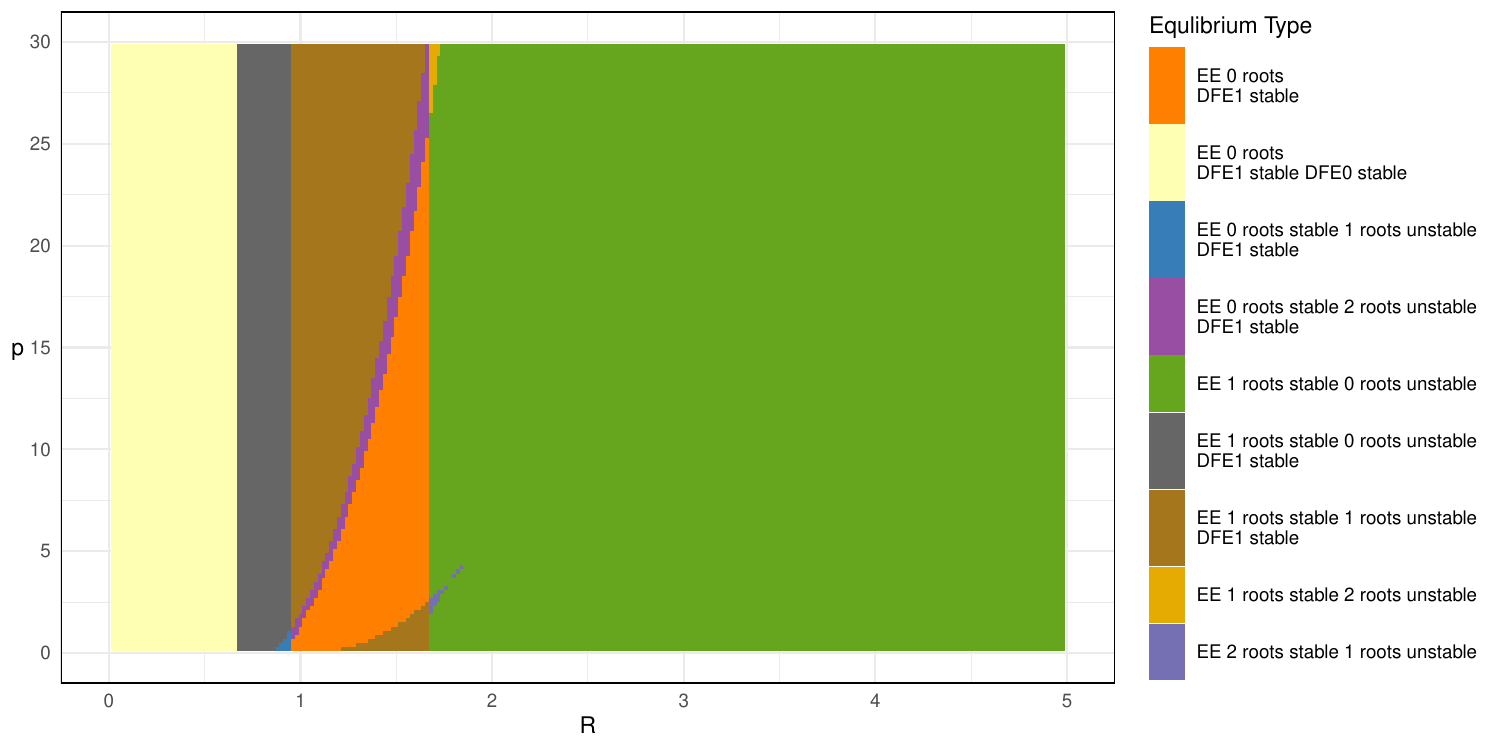}
        \caption{$\delta = 1.5$}\label{subfig:Heatmap3_b}
    \end{subfigure}
    \caption{Stability domains in the $R$--$p$ plane for $\delta=1.0530242665286913$ (upper figure) and $\delta=1.5$ (bottom figure) with $\alpha=0.6$. The colors indicate the stability domains of the different equilibria. The point labeled \emph{orbit} depicts the location of parameter regime of trajectories in Figures~\ref{F:Orbit} and~\ref{F:Orbit-zoom}. The acronyms in the legend translate to Endemic Equilibrium (EE), Disease Free Equilibrium 1 (DFE$_1$) and Disease Free Equilibrium 2 (DFE$_2$) from Lemmata~\ref{L:DFE:a}. We can see very different characteristics of dynamics between figures. Colors mostly overlap, however their sizes alter greatly. The purple region (DFE$_1$ stable) appears when going from $\delta=1.0530242665286913$ to $\delta=1.5$.}\label{F:Heatmap3}
\end{figure}

Analyzing the stability of these potential additional endemic equilibria requires discussing the sign of the trace of the Jacobian $\tr J = -(Rv+1)+J_{22}$ according to \eqref{E:Jac:szv}. Substituting $z=1/R$ and $v=\frac{1-s}{Rs}$ into the trace, we arrive at the condition
\begin{equation*}
    p A(\sigma) < R^2(1-\alpha)^2 \frac{2-s}{s}
\end{equation*}
where $\sigma=Rs$ and $A(\sigma)=-(\alpha^2+4\alpha+1)\sigma^2 + 6(\alpha+1)\sigma -6$. Analogously, the determinant can be written as $\det J=\frac{s-1}{R^3s^2(1-\alpha)^2} B(\sigma)$, where
\begin{equation*}
    B(\sigma) = -3\alpha(1+\alpha)p\sigma^4 + 2(1+4\alpha+\alpha^2)p\sigma ^3 -3(1+\alpha)p\sigma^2 - (1-\alpha)^2(1+\alpha)R^2\sigma -2(1-\alpha)^2R^3\;.
\end{equation*}
Unlike in the simple situation for $\alpha=0$, where the trace was always negative, we can now find parameter combinations, such that the trace condition fails. However, an explicit description of these combinations is rather lengthy. Therefore, we provide numerical results and heatmaps depicting the stability regions in the $Rp$--plane for different values of $\alpha$ and $\delta$, see Figures~\ref{F:Heatmap2} and~\ref{F:Heatmap3}.

\begin{rem}
In the limiting case of $p=0$, i.e.~susceptibles do not change their opinion, the only endemic equilibrium is given by $s=\frac{2}{(1+\alpha)R}$, $z=\frac{1}{R}$ and $v=\frac{(1+\alpha)R-2}{2R}$. It exists, if $(1+\alpha)R>2$. In this situation the Jacobian~\eqref{E:Jac:szv} reads as
\begin{equation*}
    J = \lpm -\frac{1+\alpha}{2}R & 0 & - \frac{2}{1+\alpha} \\
        -(1+\alpha) & 1-\frac{1+\alpha}{2}R & -2 \\
        0 & \frac{1+\alpha}{2}R-1 & 0 \rpm\;.
\end{equation*}
The trace equals $\tr J = 1-(1+\alpha)R<0$, the determinant satisfies $\det J=-\frac{1}{2} \lsb (1+\alpha)R-2\rsb^2<0$, and the third stability condition yields
\begin{align*}
    \tr J \cdot S_2-\det J &= \lsb (1+\alpha)R-2\rsb \cdot \lsb \tr J\cdot \left(\frac{1+\alpha}{4}R+1\right) - 1+\frac{1+\alpha}{2}R\rsb\\
        &= -\frac{(1+\alpha)R}{4}\lsb (1+\alpha)R-2\rsb \cdot \lsb 1+(1+\alpha) R\rsb <0\;.
\end{align*}
showing that this endemic equilibrium is always asymptotically stable.
\end{rem}

Last, we shortly comment on the general model~\eqref{E:SIsystem} including $\delta\neq 1$. For the disease free equilibria we obtain

\begin{lem}\label{L:DFE:d}
In the general case $\delta>0$, the following holds
\begin{enumerate}
\item The disease free equilibrium $\text{DFE}_0=(0,1,0)$ is asymptotically stable, if and only if $R < \frac{1}{\delta}$. 
\item The disease free equilibrium $\text{DFE}_1=(1,0,0)$ is asymptotically stable, if and only if $R<\frac{1}{\alpha}$.  
\item The balanced disease free equilibrium $\text{DFE}_b=(\frac12, \frac12, 0)$ is unstable for all $R>0$.
\end{enumerate}
\end{lem}

When considering endemic equilibria, we proceed as before. Stationarity of the ODEs for $y=I_b$ and $v=I_a+\alpha I_b$, yields the non--trivial equilibrium values $y=\frac{1-\alpha Rx}{\delta R}$  and $v=\frac{1-x-y}{R(x+\delta y)}$. Inserting these values into the ODE for $x=S_a$, we obtain a forth order equation
\begin{align*}
    \phi(x) & = c_4 x^4 + c_3 x^3 + c_2 x^2 + c_1 x + c_0 = 0
\intertext{to be solved for $x$, where}
    c_4 &= -R^3 \delta^2 p \lb 1 -\alpha \rb \lb \alpha +\delta \rb\;. \\
    c_3 &= -R^2 p \lb 3 \alpha^2 + 2\alpha \delta - 2\alpha -\delta \rb\;, \\
    c_2 &=  -R \lsb R^2 \alpha^3 \delta -R^2 \delta (\delta -1) \alpha^2 +
        \lb-R^2 \delta^2 - p\delta +p\rb \alpha +3 p\delta \rsb\;, \\
    c_1 &= R^3 \alpha^3 \delta + R^2 \lb R \delta - \delta + 1\rb \alpha^2
        - 2 R^2 \alpha \delta + p\;, \\
    c_0 &= R\alpha (1-R\alpha)\;.
\end{align*}
Note, that the leading coefficient $c_4$ of $\phi$ is always negative.

The real, positive roots $x$ of $\phi$ yield candidates for endemic equilibria. However, for arbitrary parameter values of $R,\alpha, p$ and $\delta$, analytical expressions for these equilibria are not helpful and we present numerical results in Section~\ref{Sec:numerical}.

\section{Numerical investigation of the 
equilibria}\label{Sec:numerical}
Numerical methods were implemented to analyze the dynamics of the system's attractors and basins of attraction. We run simulations using the Julia language~\cite{bezanson_julia_2017} version 1.9.4. To resolve systems numerically we use \textit{DynamicalSystems}~\cite{datseris_dynamicalsystemsjl_2018, datseris_nonlinear_2022} and \textit{OrdinaryDiffEq}~\cite{rackauckas_differentialequationsjl_2017}. The analyses were conducted for a non--trivial system with $R \geq 2.0 > \frac{2.0}{1+\alpha}$ where interesting dynamics emerge. All the trajectories shown below are generated through grid sampling of the initial conditions using a $20\times20\times20$ resolution, yielding $8,000$ initial points.

Figure~\ref{F:Dyn2} illustrates the system with parameters 
$R=2.0$, $\alpha=0.25$, $p=4.0$, and $\delta=1.0$, exhibiting three equilibria. Among these, the equilibrium EE$_1$ and the disease-free equilibrium (DFE) are stable, while EE$_b$ is unstable. Most trajectories converge toward the DFE, whereas those approaching EE$_1$ display oscillatory behavior. The unstable equilibrium EE$_b$ lies on the heteroclinic orbit, serving as the separatrix point that divides the basins of attraction of the two stable equilibria.

Figure~\ref{F:Dyn3} presents the same system ($R=2.0$, $\alpha=0.25$, $p=4.0$, $\delta=1.0$) in two dimensions, using as coordinates $s=Sa+Sb$ and time. This representation provides a clearer visualization of the underlying dynamics. The trajectories associated with EE$_1$ and the DFE are again evident; however, the basins of attraction display a discontinuous pattern as $s$ increases. For low values of $s$, the DFE acts as the primary attractor. As $s$ increases to approximately $0.7<s<0.8$, EE$_1$ becomes the dominant attractor. Beyond this region, the system dynamics re--switch, and the DFE again becomes the attractor near the location of EE$_b$, which corresponds to the origin of the heteroclinic orbit trajectories.

\begin{figure}[hbt]
    \centering
    \includegraphics[width=.6\textwidth]{plots/trajectories/2opinion-single-3d-R=2.0-alpha=0.25-p=4.0.png}
    \caption{3D trajectories with $R=2.0$, $\alpha=0.25$, $p=4.0$, $\delta=1.0$. The points indicate equilibria and the lines represent respective trajectories that converge to the equilibria. Colors indicate individual equilibria and their converging trajectories: Balanced Endemic Equilibrium (EE$_b$), First Endemic Equilibrium (EE$_1$), Disease Free Equilibrium (DFE). The heteroclinic orbit is depicted by the black line that originates from EE$_b$ that is an unstable equilibrium in this case. EE$_1$ and DFE are stable equilibria with most of the trajectories converging to the DFE. EE$_1$ trajectories spiral inwards towards the equilibrium.}
    \label{F:Dyn2}
\end{figure}
\begin{figure}[hbt]
    \centering
    \includegraphics[width=.6\textwidth]{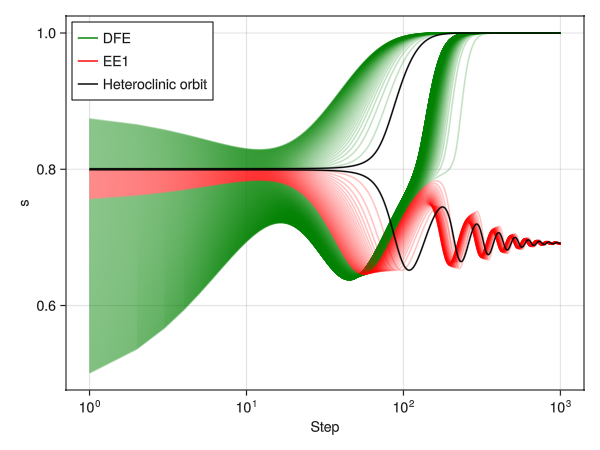}
    \caption{2D trajectories with the same parameters as in Figure~\ref{F:Dyn2}; $R=2.0$, $\alpha=0.25$, $p=4.0$, $\delta=1.0$ and in different coordinates of $s=S_a+S_b$ in linear scale and time in log scale. The $I_a$ and $I_b$ variables are set to the EE$_b$ values so the EE$_b$ lies on the sampled manifold. The lines represent respective trajectories that converge to the equilibria. Colors indicate individual equilibria and their converging trajectories: First Endemic Equilibrium (EE$_1$), Disease Free Equilibrium (DFE). The heteroclinic orbit is depicted by the black line that originates from EE$_b$ which is an unstable equilibrium in this case. EE$_1$ and DFE are stable equilibria with most of the trajectories converging to the DFE. The EE$_1$ trajectories oscillate towards the equilibrium. The attraction points of the initial conditions exhibit non-continuous basins of attraction with increasing $s$. The DFE is the attraction point for low $s$. EE$_1$ becomes the attractor somewhere between $s=0.7$ and $s=0.8$. The dynamics re-switche to the DFE as the attractor in the EE$_b$ location that is the origin of heteroclinic orbit.}
    \label{F:Dyn3}
\end{figure}

\begin{figure}[hbt]
    \centering
    \includegraphics[width=.6\textwidth]{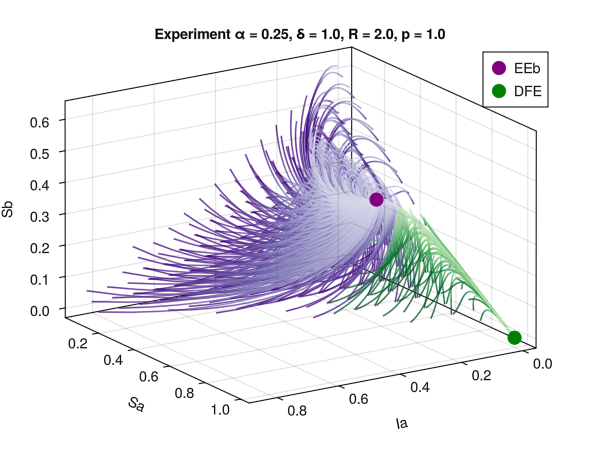}
    \caption{3D trajectories with $R=2.0$, $\alpha=0.25$, $p=1.0$, $\delta=1.0$. The points indicate equilibria and lines represent respective trajectories that converge to the points. Colors indicate individual equilibria and their converging trajectories: Balanced Endemic Equilibrium (EE$_b$), Disease Free Equilibrium (DFE). EE$_b$ and DFE are stable equilibria with most of the trajectories converging to the EE$_b$.}
    \label{F:Dyn4}
\end{figure}

Figure~\ref{F:Dyn4} depicts the system with parameters $R=2.0$, $\alpha=0.25$, $p=1.0$, $\delta=1.0$, which exhibits two stable equilibria: EE$_b$ and the disease--free equilibrium (DFE). Most trajectories converge toward EE$_b$. Because EE$_b$ is stable, it does not act as a separatrix, and no heteroclinic orbit is present. Moreover, no oscillatory behavior is observed in this case. The resulting dynamics differ markedly from those illustrated in Figure~\ref{F:Dyn2}.

To examine the dynamics of the basins of attraction with changing parameters we use the Muller plots with the Relative Equilibrium Attractivity (REA) depicted. The REA is defined by the fraction of trajectories converging to the equilibrium among all the feasible trajectories. To visualize the expected behavior of the system we compute the expected number of infected $\mathbb{E}[I_a + I_b]$ defined as the weighted mean of the number of infected in the equilibria and their respective Relative Equilibrium Attractivity.

\begin{figure}[hbt]
    \centering
\includegraphics[width=0.6\linewidth]{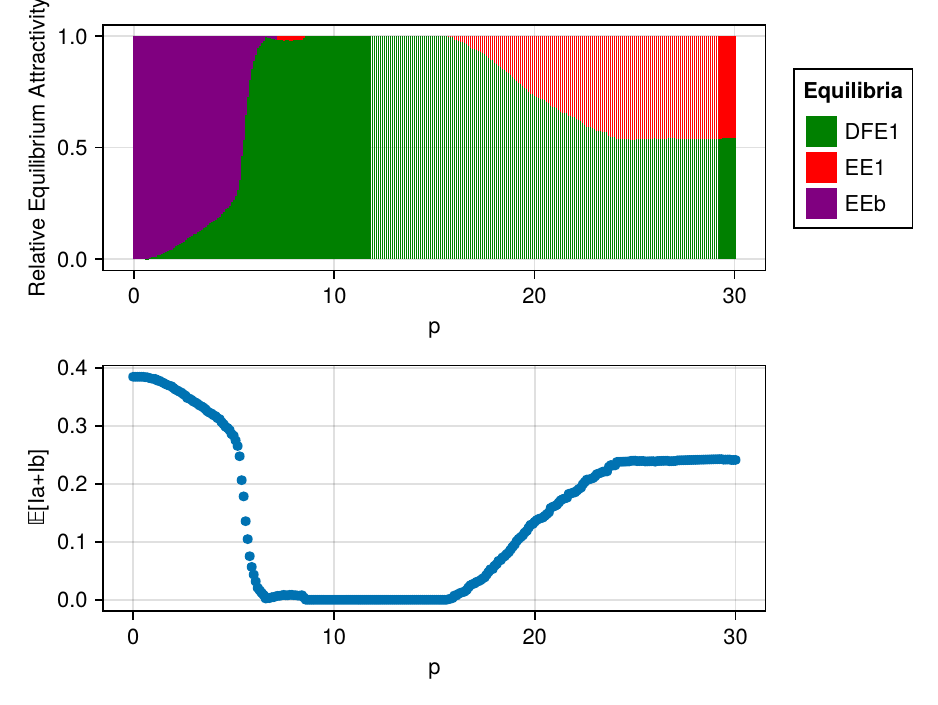}
\caption{\label{F:Infect:vs:p}Relative Equilibrium Attractivity as a size of its basin of attraction (top) and $\mathbb{E}[I_a+I_b]$ as expected number of infected (bottom) vs. $p$ for $R=2.5$, $\alpha=0.3$ and $\delta=1.0$. In the top Muller Plot colors indicate individual equilibria and their converging basins of attration: Disease Free Equilibrium 1 (DFE$_1$), Endemic Equilibrium 1 (EE$_1$), Balanced Endemic Equilibrium (EE$_b$). Relative Equilibrium Attractivity is depicted by the fraction of basins of attraction of respective equilibria in the initial condition space. The expected number of infected is a weighted mean of the number of infected in the equilibria and their respective Relative Equilibrium Attractivity.}
\end{figure}

Figure~\ref{F:Infect:vs:p} illustrates the dynamics of the REA as the parameter $p$ increases, with $R=2.5$, $\alpha=0.3$ and $\delta=1.0$ held constant. For low values of $p$, the dynamics of opinion change are too slow to effectively mitigate the epidemic. As $p$ increases, however, the basin of attraction of the disease-free equilibrium (DFE$_1$) expands, eventually encompassing the entire feasible initial condition space. In this regime, the expected number of infected individuals approaches zero.

For even larger values of $p$, the opinion change process becomes so rapid that the system no longer fully converges to the protective DFE$_1$ state. Instead, it diverges from complete epidemic mitigation toward a mixed regime characterized by the coexistence of DFE$_1$ and EE$_1$ attractors. At high $p$, the system reaches a saturated, polarized state in which approximately half of the initial conditions converge to DFE$_1$ and the other half to EE$_1$.

\begin{figure}[hbt]
    \centering
\includegraphics[width=0.6\linewidth]{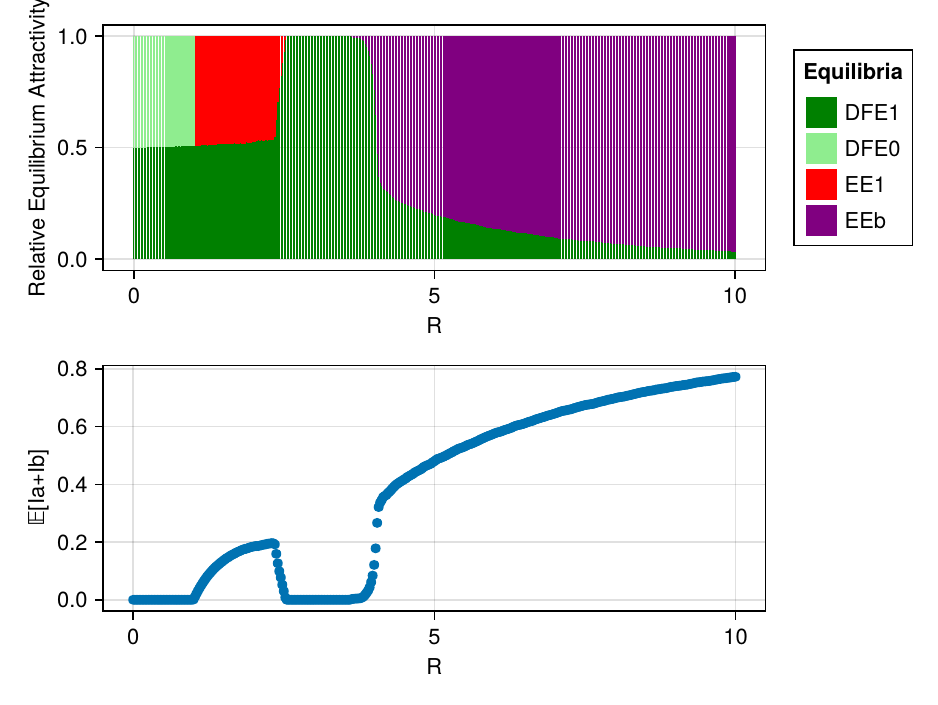}
\caption{\label{F:Infect:vs:R}Relative Equilibrium Attractivity as a size of its basin of attraction (top) and $\mathbb{E}[I_a+I_b]$ as expected number of infected (bottom) vs. $R$ for $\alpha=0.0$, $p=11.5$ and $\delta=1.0$. In top Muller Plot colors indicate individual equilibria and their converging basins of attraction: Disease Free Equilibrium 1 (DFE$_1$), Disease Free Equilibrium 0 (DFE$_0$), Endemic Equilibrium 1 (EE$_1$), Balanced Endemic Equilibrium (EE$_b$). Relative Equilibrium Attractivity is depicted by the fraction of basins of attraction of respective equilibria in the initial condition space. The expected number of infected is a weighted mean of the number of infected in the equilibria and their respective Equilibrium Density.}
\end{figure}

Figure~\ref{F:Infect:vs:R} presents a similar analysis of the REA and the expected number of infected individuals as the parameter $R$ increases, with $\alpha=0.0$, $p=11.5$, and $\delta=1.0$ held constant. For low values of $R$, the epidemic is effectively mitigated, and the system converges to either 
DFE$_0$ or DFE$_1$. When $R>1.0$, two equilibria, DFE$_1$ and EE$_1$, coexist, and the dynamics resemble those observed in Figure~\ref{F:Infect:vs:p} for $p>25.0$, where the rate of opinion change is too high to sustain full epidemic mitigation.

As $R$ increases beyond $2.5$, the system consistently mitigates the epidemic, regardless of the initial conditions. For even larger values of $R$, a stable equilibrium EE$_b$ emerges and gradually becomes the dominant attractor. In this regime, the opinion change mechanism is generally too slow to counteract the epidemic spread, leading to dynamics predominantly driven by infection processes.

Interestingly, system with $\alpha=0.6$, $\delta= 1.0530242665286913$, $R=2.0$ and $p=11.5$ exhibits orbital trajectories with complex eigenvalues such that the real part is negative and close to zero up to numerical precision. The numerical values with precision up to 3 decimals are given by
\begin{equation*}
    [-1.123,\; -1.57\cdot 10^{-15} - 0.651i,\; -1.57\cdot 10^{-15}+0.651i]\;.    
\end{equation*}

Furthermore, if we increase the parameter to $\delta= 1.0530242665286915$ the real part of complex eigenvalues is positive:
\begin{equation*}
[-1.123,\; 1.18 \cdot 10^{-16} - 0.651i,\; 1.18\cdot 10^{-16} + 0.651i]
\end{equation*}
This suggests that somewhere between above delta values the real part should be equal to zero and system will exhibit a limit cycle. However, we could not find the exact spot numerically.

\begin{figure}[hbt]
    \centering
\includegraphics[width=0.6\linewidth]{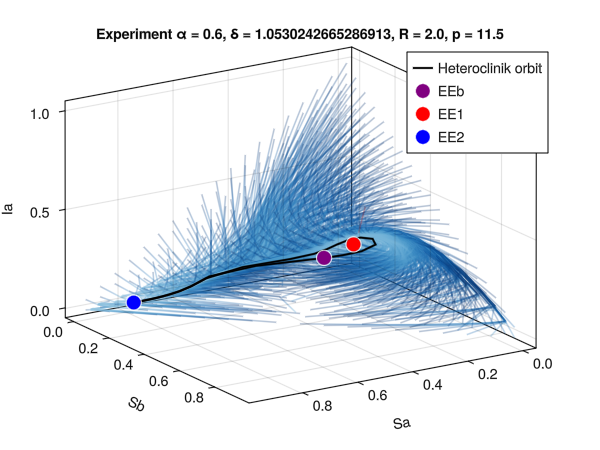}
\caption{\label{F:Orbit}3D trajectories with $R=2.0$, $\alpha=0.6$, $p=11.5$, $\delta=1.0530242665286913$. The points indicate equilibria and lines represent respective trajectories that converge to the points. Colors indicate individual equilibria and their converging trajectories: Balanced Endemic Equilibrium (EE$_b$), First Endemic Equilibrium (EE$_1$), Second Endemic Equilibrium (EE$_2$). Heteroclinic orbit is depicted by black line that originates from EE$_b$ that is an unstable equilibrium in this case. Both EE$_1$ and EE$_2$ equilibria are stable with all of the trajectories converging to it. Many EE$_2$ trajectories and heteroclinic orbit make turn around EE$_1$ equilibrium to finally converge to EE$_2$.}
\end{figure}

\begin{figure}[hbt]
    \centering
\includegraphics[width=0.6\linewidth]{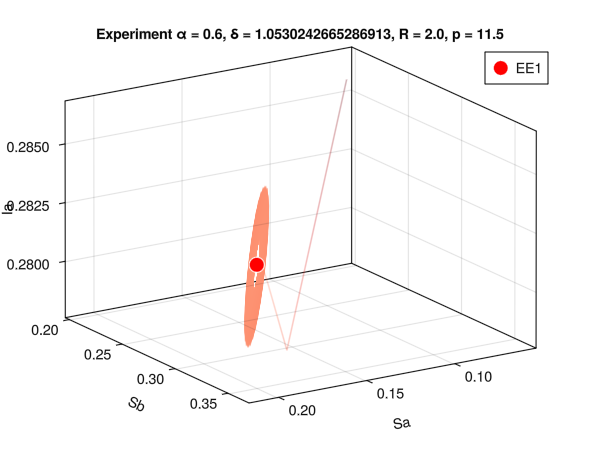}
\caption{\label{F:Orbit-zoom}Zoom on 3D trajectories of EE$_1$ equilibrium from Figure~\ref{F:Orbit} with $R=2.0$, $\alpha=0.6$, $p=11.5$, $\delta=1.0530242665286913$. The point indicate stable EE$_1$ equilibrium and line represent respective trajectory that converges to the point. EE$_1$ trajectory spirals rapidly around EE$_1$ equilibrium and would finally converge to it. However due to feasibility the simulation stopped after $20,000$ steps. Due to small basin of attraction the resolution was increased to $125,000$ initial points.}
\end{figure}

In Figure~\ref{F:Orbit}, the overall dynamics of the system with $\alpha=0.6$, $\delta= 1.0530242665286913$, $R=2.0$ and $p=11.5$ are illustrated. From a broad perspective, no particularly unusual behavior is immediately apparent. The system possesses two stable equilibria, denoted as EE$_1$ and EE$_2$, while the equilibrium point EE$_b$ is unstable. To provide a clearer visualization of the orbit, Figure~\ref{F:Orbit-zoom} focuses exclusively on trajectories associated with EE$_1$. It can be observed that the trajectory of $20,000$ steps converges extremely slowly toward the EE$_1$ equilibrium. 

\section{Discussion and Outlook}

The presented model couples epidemic dynamics, i.e.~transmission of a pathogen and social dynamics, i.e.~the exchange of opinions. These opinions have an influence on the transmission dynamics by altering the risk of transmission and infection. Besides the epidemiological parameters $R,\alpha$ and $\delta$, our model includes the rate of opinion change $p$ as a social parameter. One can associate $p$ with the communication speed inside the population; faster communication will typically lead to an increased rate of opinion change. From a public health point of view, it is interesting to analyze, how the total number of infected individuals $I_a+I_b$ depends on the rate of opinion change $p$. Figure~\ref{F:Infect:vs:p} shows the expected value of $I_a+I_b$ versus the rate of opinion change $p$.
For small values of $p$, opinion dynamics evolve too slowly to meaningfully curb the epidemic. As $p$ increases, the basin of attraction for the disease-free equilibrium (DFE$_1$) grows and eventually covers the entire feasible space of initial conditions, driving the expected number of infected individuals toward zero (bottom sub-figure). When 
$p$ becomes even larger, however, opinion change occurs so quickly that the system no longer fully settles into the protective DFE$_1$ state. Instead, it shifts into a mixed regime in which both DFE$_1$ and EE$_1$ act as attractors. At high $p$, the system enters a saturated, polarized state where roughly half of the initial conditions converge to DFE$_1$ and the other half to EE$_1$.

The Figure~\ref{F:Infect:vs:R} shows the expected value of $I_a + I_b$ versus the rate of opinion change $R$. For small values of $R$, the epidemic is effectively controlled, and the system settles into either DFE$_0$ or DFE$_1$. When $R > 1.0$, the equilibria DFE$_1$ and EE$_1$ coexist, producing dynamics similar to those in Figure~\ref{F:Infect:vs:p} for $p > 25.0$, where the rate of opinion change becomes too rapid to maintain full epidemic suppression.

As $R$ exceeds $2.5$, the system reliably mitigates the epidemic regardless of the starting conditions. For still larger values of $R$, a stable equilibrium EE$_b$ appears and gradually becomes the dominant attractor. In this regime, opinion dynamics evolve too slowly to counter the spread of infection, resulting in behavior primarily driven by epidemic processes.

The results demonstrate the significant role of spread of public opinion in the dynamics of epidemic transmission. The relationship between the basic reproduction number, 
$R$, and the rate of opinion change, 
$p$, has a profound impact on the course of an epidemic. There exists an optimal, or balanced, range for these parameters in which epidemic control and potential extinction are achievable. Specifically, neither 
$R$ nor $p$ should be excessively high or low in relation to each other, as imbalances in either parameter can lead to suboptimal outcomes.

While the spread of opinion can have a mitigating effect on epidemic transmission under certain conditions, this effect is not unlimited. It is constrained by the underlying dynamics of the epidemic and the rate at which opinion shifts occur. If the rate of opinion change, 
$p$, becomes too rapid or extreme, the influence of opinion on epidemic spread can backfire, leading to a counterproductive effect. In such cases, rapid shifts in public opinion may result in a prolonged endemic state, where the epidemic persists at a relatively stable, but high, level of infection. This situation arises for a large portion of initial conditions, particularly when the opinion change rate is either too fast or too erratic. Consequently, the expected number of infections may increase substantially compared to scenarios with more moderate opinion shifts.

In summary, while public opinion can play a role in mitigating epidemic spread, its effectiveness is contingent upon maintaining a delicate balance between 
$R$ and $p$. If either factor is misaligned, the result can be a counterproductive scenario where the epidemic is not controlled, and in some cases, may even intensify.

As with any abstraction of real-world processes, our model is necessarily limited in scope, and several avenues for extension remain open.

A natural next step would be a comparison of the model’s quantitative predictions with empirical data. Such validation would be highly valuable, especially if conducted over time to capture the coevolution of opinions and epidemic dynamics. Unfortunately, to the best of the authors’ knowledge, no longitudinal datasets currently exist that simultaneously track epidemiological variables and population-level opinion shifts with sufficient resolution.

In this work, we focused on a single opinion-spread mechanism, the 
$q$-voter model. Given the strong influence that opinion dynamics exert on epidemic outcomes in our results, it would be worthwhile to explore how alternative communication and persuasion mechanisms shape epidemic trajectories, and to identify which behavioral regimes yield the most effective mitigation. Many established models could be considered, including but not limited to the bounded-confidence frameworks, the Sznajd model, and majority-rule or kinetic-exchange models. A systematic survey of such mechanisms may reveal qualitatively different regimes of epidemic suppression, persistence, or amplification.

Our framework is also constrained by its formulation as a compartmental ODE model, which relies on the assumption of homogeneous mixing within and between compartments. Real-world social and epidemiological processes are considerably more heterogeneous and structured. Extensions could therefore incorporate non--homogeneous populations--e.g., stratification by age, contact patterns, or communication behavior--to better reflect demographic variability in both disease spread and opinion adoption. 

Finally, a promising direction is the explicit inclusion of mass media and social media as external drivers of opinion dynamics. These entities facilitate rapid propagation of information yet do not participate in the epidemic process themselves. Introducing such media as exogenous fields or network layers could improve the realism of opinion--epidemic coevolution and yield more nuanced insights into how public messaging shapes collective behavior and health outcomes.

\end{document}